\documentclass[a4paper]{article}

\usepackage{INTERSPEECH2021}
\usepackage{amsmath,graphicx, hyperref,xcolor,acronym, comment}
\usepackage{amsmath,graphicx, hyperref,xcolor,colortbl,subfig, comment}

\title{INTERSPEECH 2021 Deep Noise Suppression Challenge}
\name{Chandan K A Reddy$^1$, Harishchandra Dubey$^1$, Kazuhito Koishida$^1$, Arun Nair$^2$, Vishak Gopal$^1$, Ross Cutler$^1$, Sebastian Braun$^1$, Hannes Gamper$^1$, Robert Aichner$^1$, Sriram Srinivasan$^1$}
\address{
  $^1$Microsoft Corporation\\
  $^2$Johns Hopkins University}
  
\email{chkarada@microsoft.com, firstname.lastname@microsoft.com}

\begin{document}

\maketitle
\begin{abstract}
  The Deep Noise Suppression (DNS) challenge was designed to unify the research efforts in the area of noise suppression targeted for human perception. We recently organized a DNS challenge special session at INTERSPEECH 2020 and ICASSP 2021. We open-sourced training and test datasets for the wideband scenario along with a subjective evaluation framework based on ITU-T standard P.808, which was used to evaluate participants of the challenge. Many researchers from academia and industry made significant contributions to push the field forward, yet even the best noise suppressor was far from achieving superior speech quality in challenging scenarios. In this version of the challenge organized at INTERSPEECH 2021, we expanded our training and test datasets to accommodate fullband scenarios and challenging test conditions. We used ITU-T P.835 to evaluate the challenge winners as it gives additional information about the quality of processed speech and residual noise. The two tracks in this challenge focused on real-time denoising for (i) wideband, and (ii) fullband scenarios. We also made available a reliable non-intrusive objective speech quality metric for wideband called DNSMOS for the participants to use during their development phase. 
\end{abstract}
\vspace{2mm}

\section{Introduction}

With the explosion in the number of people working remotely due to the pandemic, there has been a surge in the demand for reliable collaboration and real-time communication tools. Excellent speech quality in our audio calls is a need during these times as we try to stay connected and collaborate with people every day. We are easily exposed to a variety of background noises such as a leaf blower, washing machine, dog barking, a baby crying, kitchen noises, etc. Background noise significantly degrades the quality and intelligibility of the perceived speech leading to fatigue. Background noise poses a challenge in other applications such as hearing aids and smart devices as well. 

Real-time Speech Enhancement (SE) for perceptual quality is a decades-old classical problem and researchers have proposed numerous solutions~\cite{malah, 8031044}. In recent years, learning-based approaches have shown promising results~\cite{8281993, choi2020phase, koyama2020exploring}. The Deep Noise Suppression (DNS) Challenge organized at INTERSPEECH 2020 \cite{reddy2020interspeech} and ICASSP 2021 \cite{reddy2020icassp} showed great progress, while also indicating that we are still about 1.6 Differential Mean Opinion Score (DMOS) away from the ideal Mean Opinion Score (MOS) of 5 when tested on the challenge test set, which was reasonably representative of realistic scenarios. The DNS Challenge is the first contest that we are aware of using the subjective evaluation to benchmark SE methods using a realistic noisy test set \cite{reddy2020interspeech}.    

We open-sourced a large dataset for INTERSPEECH 2020 and ICASSP 2021 DNS challenge\footnote{\url{https://github.com/microsoft/DNS-Challenge}}. For ease of reference, we will call the INTERSPEECH 2021 challenge DNS Challenge 3, ICASSP 2021 challenge DNS Challenge 2, and the INTERSPEECH 2020 challenge DNS Challenge 1. The DNS Challenge 3 was focused on real-time denoising similar to track 1 of both the DNS challenges 1 and 2. We had 2 tracks in DNS challenge 3 for wideband (sampling rate = 16000 Hz) and fullband (sampling rate = 48000 Hz) scenarios. The datasets include over 760 hours of clean speech including singing voice, emotion data, and non-English languages. Noise data in the training set remains the same as DNS Challenge 2. Both clean speech and noise are made publicly available for both wide and fullband scenarios. We provide over 118,000 room impulse responses (RIR), which includes real and synthetic RIRs from public datasets for wideband. We provide acoustic parameters: Reverberation time (T60) and Clarity (C50) for read clean speech and RIR sample. The test set includes a variety of noisy speech utterances in English and non-English in a range of reverberant and noisy scenarios. We also include emotional speech and singing in the presence of background noise. 

Unlike ITU-T P.808 that was used for DNS Challenge 1 and 2,  we used the implementation of ITU-T P.835\footnote{\url{https://github.com/microsoft/P.808}} \cite{naderi2020crowdsourcing} for the DNS Challenge 3. In addition to the overall speech quality as in P.808, P.835 provides standalone quality scores of speech and noise. The standalone ratings will help us focus on the areas that require improvement to achieve better overall speech quality. Many noise suppressors are very good in suppressing the background noise but do not improve the quality of speech, which becomes the bottleneck for improving the overall quality. The results of this challenge discussed in the later sections reflect the same. We also provided a non-intrusive objective speech quality metric for wideband scenario called DNSMOS\footnote{\url{https://github.com/microsoft/DNS-Challenge/tree/master/DNSMOS}} as an Azure service. We showed that DNSMOS is more reliable than other widely used objective metrics such as PESQ, SDR, and POLQA \cite{reddy2020dnsmos}. Also, it does not require reference clean speech and hence can work on real recordings. This paper describes the datasets, challenge results, and the learnings from the challenge in more detail.

\section{Challenge Tracks}
The following were the algorithmic and computational requirements that each participant had to satisfy to be eligible for the challenge. 
\begin{enumerate}
    \item Track 1: Real-Time Denoising track for wideband scenario
    \begin{itemize}
        \item The noise suppressor must take less than the stride time $T_s$ (in ms) to process a frame of size $T$ (in ms) on an Intel Core i5 quad-core machine clocked at 2.4 GHz or equivalent processor. For example, $T_s = T/2$ for 50\% overlap between frames. The total algorithmic latency allowed including the frame size $T$, stride time $T_s$, and any look ahead must be $\leq$ 40ms. For example, for a real-time system that receives 20ms audio chunks, if you use a frame length of 20ms with a stride of 10ms resulting in an algorithmic latency of 30ms, then you satisfy the latency requirements. If you use a frame of size 32ms with a stride of 16ms resulting in an algorithmic latency of 48ms, then your method does not satisfy the latency requirements as the total algorithmic latency exceeds 40ms. If your frame size plus stride $T_1=T+T_s$ is less than 40ms, then you can use up to $(40-T_1)$ms future information.   
    \end{itemize}
    \item Track 2: Real-Time Denoising track for fullband scenario
    \begin{itemize}
        \item Satisfy Track 1 requirements.
    \end{itemize}
\end{enumerate}

\vspace{-2mm}
\section{Training Datasets}
The goal of releasing the clean speech and noise datasets is to provide researchers with an extensive and representative dataset to train their SE models. We initially released MSSNSD \cite{reddy2019scalable} with a focus on extensibility, but the dataset lacked the diversity in speakers, emotions, languages, and noise types. We published a significantly larger and more diverse data set with configurable scripts for DNS Challenge 1 and 2 \cite{reddy2020interspeech}. Many researchers found this dataset useful to train their noise suppression models and achieved good results. However, the training and the test datasets again lacked more clips with emotions such as crying, yelling, laughter or singing. Also, the dataset only included the clips in the English language. For DNS Challenge 3, we added speech clips with other emotions and included about 10 non-English languages. The clean speech in the training set resulted in a total of 760.53 hours: read speech (562.72 hours), singing voice (8.80 hours), emotion data (3.6hours), Chinese mandarin data (185.41 hours). We have grown clean speech to 760.53 hours as compared to 562.72 hours in DNS Challenge 1. The details about the clean and noisy dataset are described in the following sections.

\subsection{Clean Speech}
\label{ssec:cleanspeech}
Clean speech consists of three subsets: (i) Read speech recorded in clean conditions; (ii) Singing clean speech; (iii) Emotional clean speech; and (iv) Non-English clean speech. The first subset is derived from the public audiobooks dataset called Librivox\footnote{https://librivox.org/}. It is available under the permissive Creative Commons 4.0 license \cite{7178964}. It has recordings of volunteers reading over 10,000 public domain audiobooks in various languages, the majority of which are in English. In total, there are 11,350 speakers. Many of these recordings are of excellent speech quality, meaning that the speech was recorded using good quality microphones in silent and less reverberant environments. But there are many recordings that are of poor speech quality as well with speech distortion, background noise, and reverberation. Hence, it is important to clean the data set based on speech quality. We used the online subjective test framework ITU-T P.808 \cite{naderi2020open} to sort the book chapters by subjective quality. The audio chapters in Librivox are of variable length ranging from few seconds to several minutes. We randomly sampled 10 audio segments from each book chapter, each of 10 seconds in duration. For each clip, we had 2 ratings, and the MOS across all clips was used as the book chapter MOS. Figure 1 shows the results, which show the quality spanned from very poor to excellent quality. In total, it is 562 hours of clean speech, which was part of DNS Challenge 1.

The second subset consists of high-quality audio recordings of singing voices recorded in noise-free conditions by professional singers. This subset is derived from~\textit{VocalSet} corpus~\cite{wilkins2018vocalset} with Creative Commons Attribution 4.0 International License (CC BY 4.0). license. It has 10.1 hours of clean singing voice recorded by 20 professional singers: 9 males, and 11 females. This data was recorded on a range of vowels, a diverse set of voices on several standards and extended vocal techniques, and sung in contexts of scales, arpeggios, long tones, and excerpts. For wideband, we downsampled the mono .WAV files from 44.1 kHz to 16 kHz and added them to the clean speech corpus used by the training data synthesizer. 

The third subset consists of emotional speech recorded in noise-free conditions. This is derived from Crowd-sourced Emotional Multimodal Actors Dataset (CREMA-D)~\cite{cao2014crema} made available under the Open Database License. It consists of 7,442 audio clips from 91 actors: 48 male, and 43 female accounting for a total of 3.5 hours of audio. The age of the actors was in the range of 20 to 74 years with diverse ethnic backgrounds including African American, Asian, Caucasian, Hispanic, and Unspecified. Actors read from a pool of 12 sentences for generating this emotional speech dataset. It accounts for six emotions: Anger, Disgust, Fear, Happy, Neutral, and Sad at four intensity levels: Low, Medium, High, Unspecified. The recorded audio clips were annotated by multiple human raters in three modalities: audio, visual, and audio-visual. Categorical emotion labels and real-value emotion level values of perceived emotion were collected using crowd-sourcing from 2,443 raters. This data was provided as 16 kHz .WAV files so we added it to our wideband clean speech as it is. 

The fourth subset has a clean speech from non-English languages. It consists of both tonal and non-tonal languages including Chinese (Mandarin), German and Spanish. Mandarin data consists of OpenSLR18~\footnote{http://www.openslr.org/18/} THCHS-30~\cite{THCHS30_2015} and OpenSLR33~\footnote{http://www.openslr.org/33/} AISHELL~\cite{aishell_2017} datasets, both with the Apache 2.0 license. THCHS30 was published by the Center for Speech and Language Technology (CSLT) at Tsinghua University for speech recognition. It consists of 30+ hours of clean speech recorded at 16-bit 16 kHz in noise-free conditions. Native speakers of standard Mandarin read text prompts chosen from a list of 1000 sentences. We added the entire THCHS-30 data in our clean speech for the training set. It consisted of 40 speakers: 9 male, 31 female in the age range of 19-55 years. It has total 13,389 clean speech audio files~\cite{THCHS30_2015}. The AISHELL dataset was created by Beijing Shell Shell Technology Co. Ltd. It has clean speech recorded by 400 native speakers ( 47\% male and 53\% female) of Mandarin with different accents. The audio was recorded in noise-free conditions using high-fidelity microphones. It is provided as 16-bit 16kHz .wav files. It is one of the largest open-source Mandarin speech datasets. We added the entire AISHELL corpus with 141,600 utterances spanning 170+ hours of clean Mandarin speech to our training set.

Spanish data is 46 hours of clean speech derived from OpenSLR39, OpenSLR61, OpenSLR71, OpenSLR73, OpenSLR74 and OpenSLR75 where re-sampled all .WAV files from 48 kHz to 16 kHz to use them as wideband signals. German data is derived from four corpora namely (i) The Spoken Wikipedia Corpora~\cite{swc}, (ii) Telecooperation German Corpus for Kinect~\cite{kinect}, (iii) M-AILABS data~\cite{mailabs}, (iv) zamia-speech forschergeist corpora. Complete German data constitute 636 hours. Italian (128 hours), French (190 hours), Russian (47 hours) are taken from M-AILABS data~\cite{mailabs}. M-AILABS Speech Dataset is a  publicly available multi-lingual corpora for training speech recognition and speech synthesis systems.
\begin{figure}[!tb]
\centering
\includegraphics[width=0.8\columnwidth]{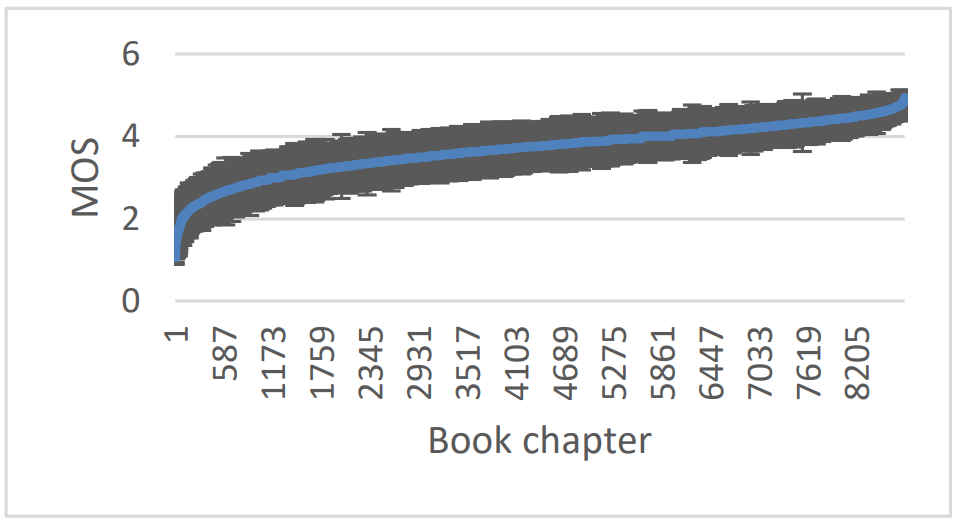}
\caption{Sorted near-end single-talk clip quality (P.808) with 95\% confidence intervals.}
\label{fig:nearend}
\end{figure}
\vspace{-1mm}
\subsection{Noise}
\label{ssec:noise}
 The noise clips were selected from Audioset \footnote{https://research.google.com/audioset/} \cite{7952261} and Freesound \footnote{https://freesound.org/}. Audioset is a collection of about 2 million human labeled 10s sound clips drawn from YouTube videos and belong to about 600 audio events. Like the Librivox data, certain audio event classes are over-represented. For example, there are over a million clips with audio classes music and speech and less than 200 clips for classes such as toothbrush, creak, etc. Approximately 42\% of the clips have a single class, but the rest may have 2 to 15 labels. Hence, we developed a sampling approach to balance the dataset in such a way that each class has at least 500 clips. We also used a speech activity detector to remove the clips with any kind of speech activity, to strictly separate speech and noise data. The resulting dataset has about 150 audio classes and 60,000 clips. We also augmented an additional 10,000 noise clips downloaded from Freesound and DEMAND databases \cite{demand}. The chosen noise types are more relevant to VoIP applications. In total, there is 181 hours of noise data. The noise files were originally of fullband, which were resampled to 16 kHz for wideband use case.
\vspace{-1mm}
\subsection{Room Impulse Responses}
We provide 3076 real and approximately 115,000 synthetic rooms impulse responses (RIRs) where we can choose either one or both types of RIRs for convolving with clean speech. Noise is then added to reverberant clean speech while DNS models are expected to take noisy reverberant speech and produce clean reverberant speech. Challenge participants can do both de-reverb and denoising with their models if they prefer. These RIRs are chosen from openSLR26~\cite{ko2017study}~\footnote{http://www.openslr.org/26/} and openSLR28~\cite{ko2017study}~\footnote{http://www.openslr.org/28/} datasets, both released with Apache 2.0 License.
\vspace{-1mm}
\subsection{Acoustic parameters}
We provide two acoustic parameters: (i) Reverberation time, T60~\cite{antsalo2001estimation} and (ii) Clarity, C50~\cite{gamper2020blind} for all audio clips in clean speech of training set. We provide T60, C50 and isReal Boolean flag for all RIRs where isReal is 1 for real RIRs and 0 for synthetic ones. The two parameters are correlated. An RIR with low C50 can be described as highly reverberant and vice versa~\cite{antsalo2001estimation,gamper2020blind}. These parameters are supposed to provide flexibility to researchers for choosing a sub-set of provided data for controlled studies. 

\section{Test set}
For DNS Challenge 3, the test set included utterances in English and non-English languages recording in the presence of a variety of background noises at different SNR, target levels, and acoustic conditions. Non-English languages included tonal languages such as Punjabi, Vietnamese, Mandarin, and Cantonese. Other Non-English languages included Spanish, German, Portuguese and French. We crowd-sourced the noisy speech collection efforts to get diversity in terms of languages, acoustic conditions, noise environments, speaker's age, ethnicity and to have gender balance. The utterances were collected at a distance of 1-5 meters from the microphone when they were not using a headphone to have more reverberation. The development and blind test sets included utterances with emotions such as laughter, crying, yelling, and surprise in the presence of background noise. This is to measure the effects of noise suppressors on human emotions. Many noise suppressors tend to be aggressive and end up suppressing low-energy emotions and sounds. A small segment of the clips includes speech in the presence of musical instruments such as guitar, piano, violin playing in the background. This is to ensure that noise suppression methods do well in the presence of musical tones overlapping with speech. We also included speech collected in the presence of stationary noise as it is the most common use case scenario. All the clips were originally collected at a sampling rate of 48 kHz and were resampled to 16 kHz.

\begin{figure}[hbt!]
 \centering
  \subfloat[Speech MOS]{\includegraphics[width=0.40\textwidth]{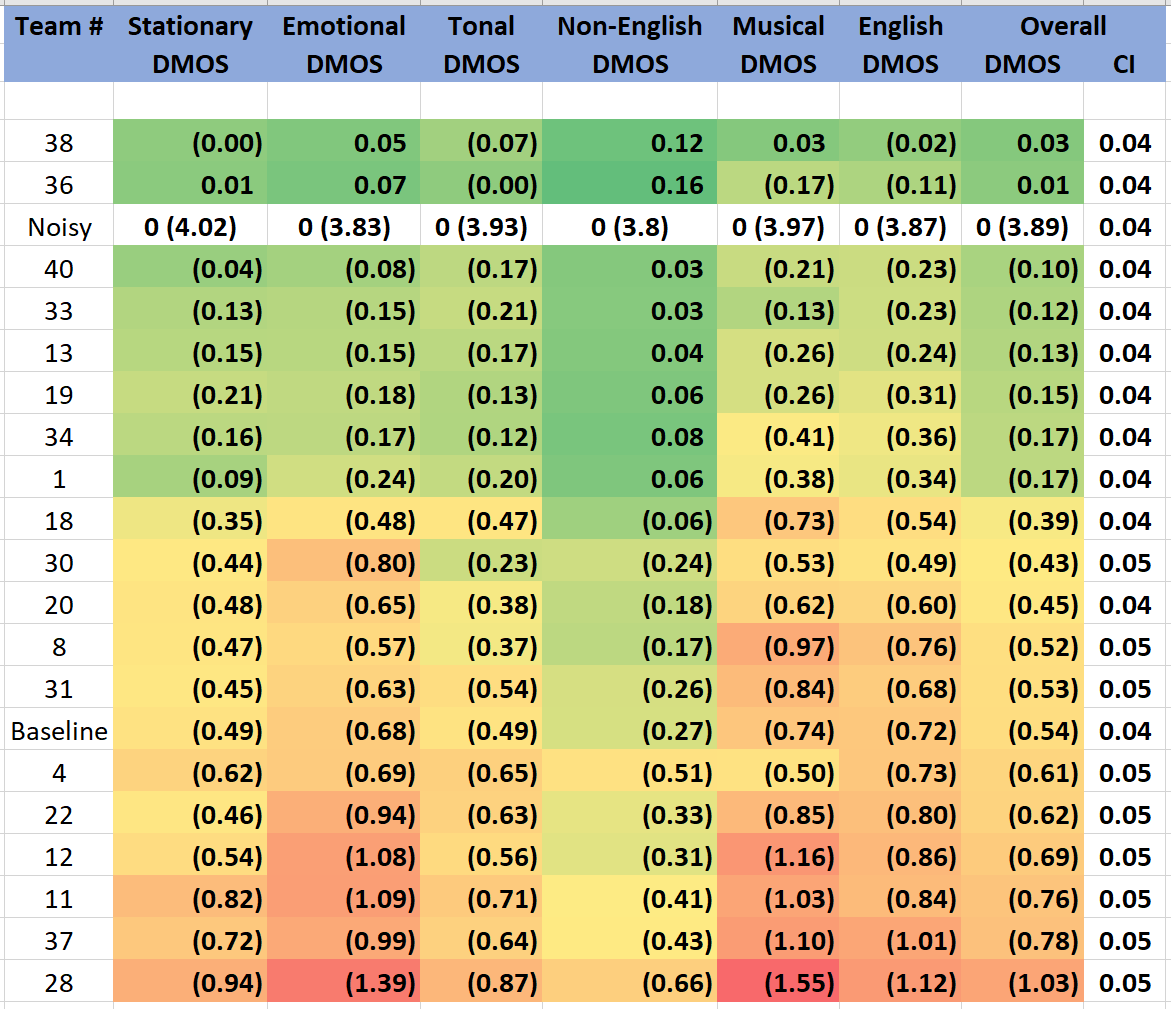}\label{fig:f1a}}
  \hfill
  \subfloat[Background Noise MOS]{\includegraphics[width=0.40\textwidth]{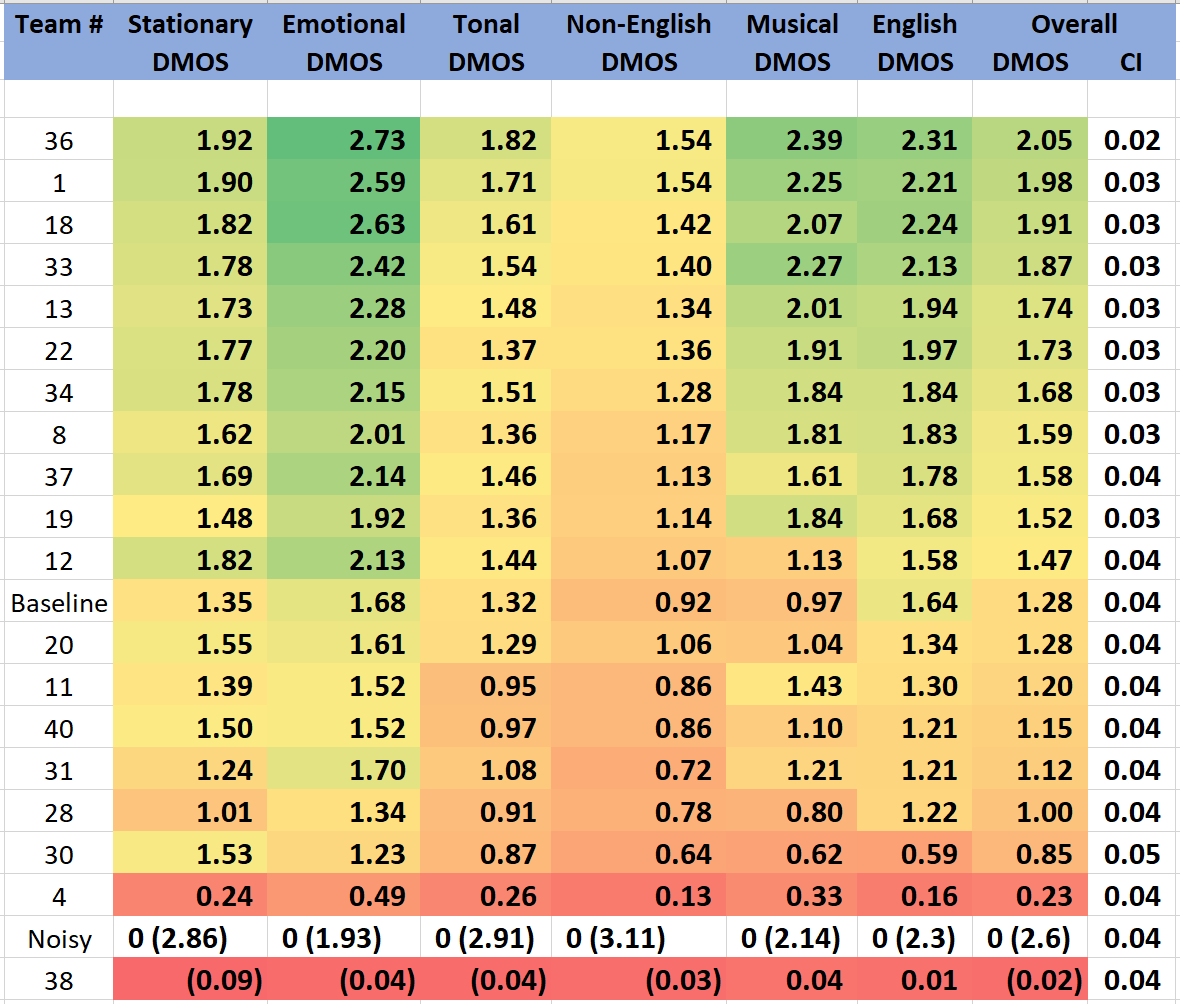}\label{fig:f1b}}
  \hfill
  \subfloat[Overall MOS]{\includegraphics[width=0.40\textwidth]{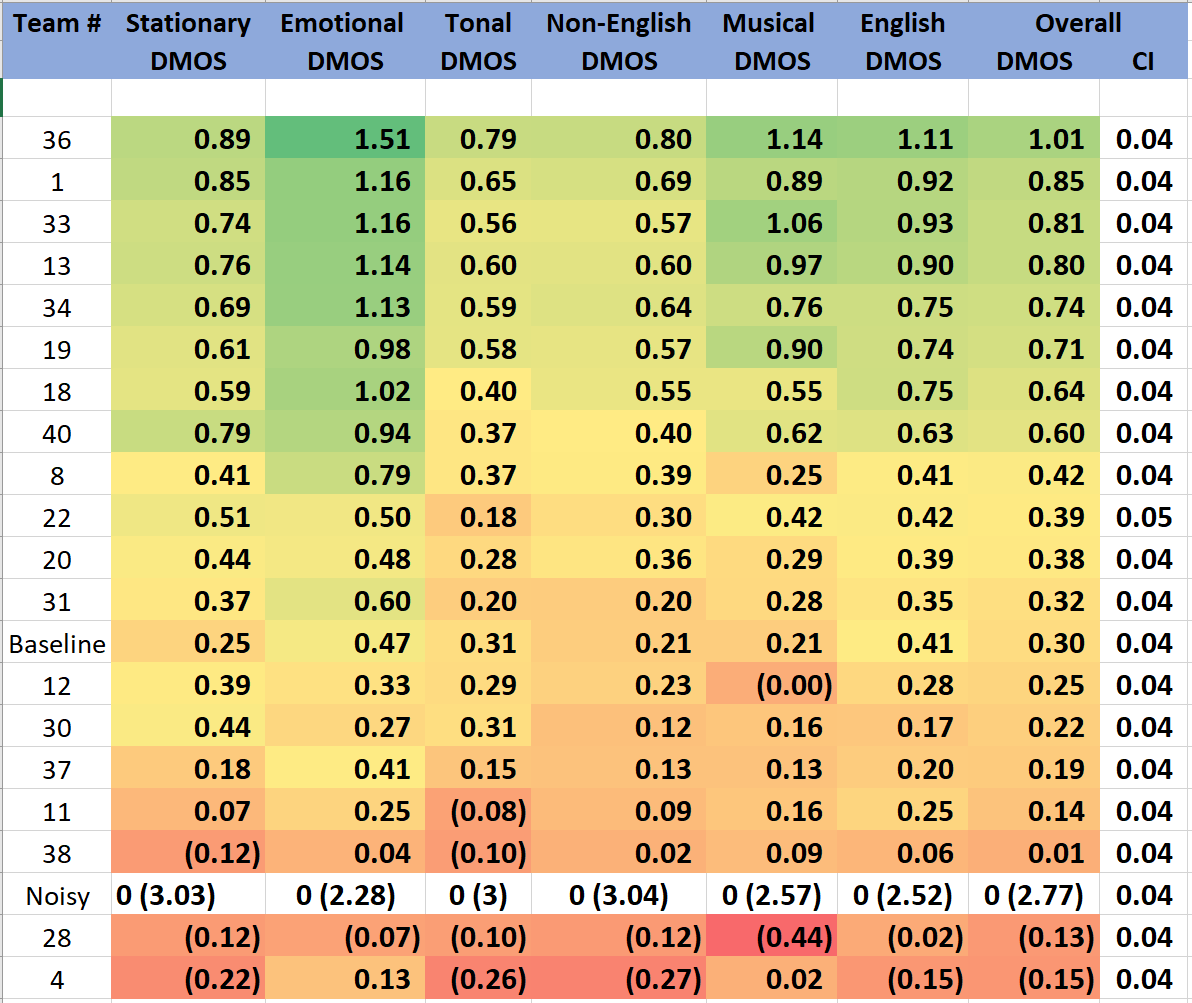}\label{fig:f1c}}
  
  \caption{Track 1 results}
\label{fig:track1}
\end{figure}

\begin{figure}[hbt!]
 \centering
  \subfloat[Speech MOS]{\includegraphics[width=0.40\textwidth]{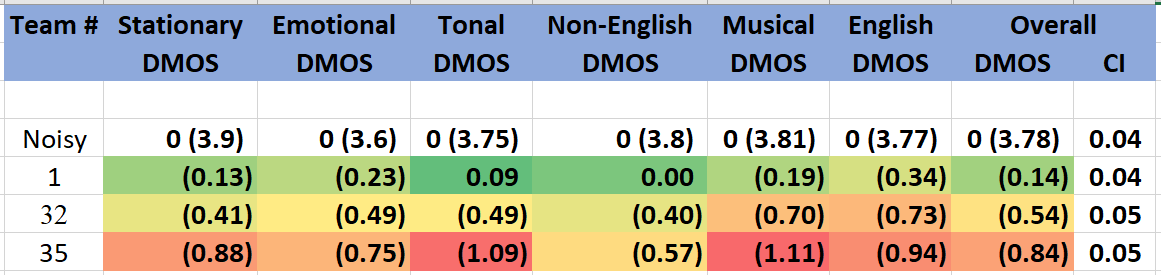}\label{fig:f3a}}
  \hfill
  \subfloat[Background Noise MOS]{\includegraphics[width=0.40\textwidth]{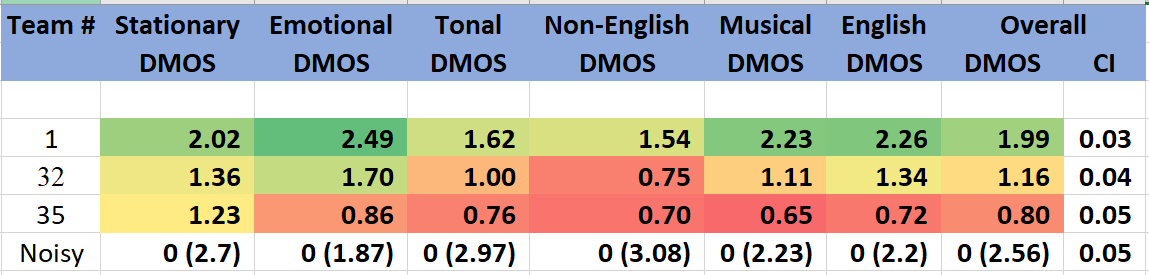}\label{fig:f3b}}
  \hfill
  \subfloat[Overall MOS]{\includegraphics[width=0.40\textwidth]{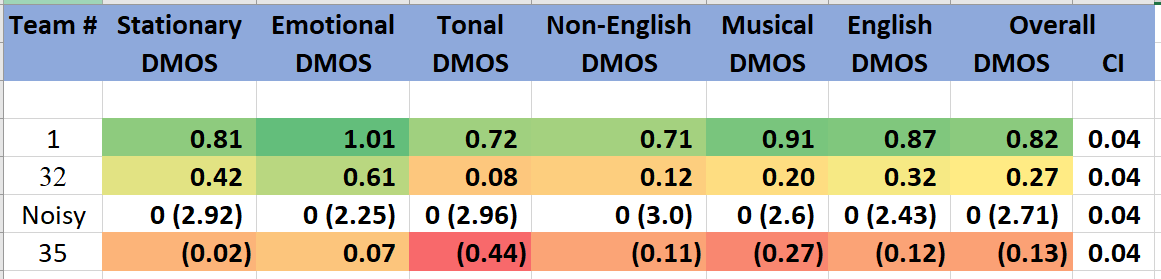}\label{fig:f3c}}
  
  \caption{Track 2 results}
\label{fig:track2}
\end{figure}

\section{Challenge Results and Key Takeaways}
\subsection{Evaluation set up and results}
The final evaluation was done on the blind test set using the crowdsourced subjective evaluation framework based on ITU-T P.835
\cite{naderi2020crowdsourcing} to determine the DNS quality. Each clip was rated by 5 qualified raters, which gave the maximum 95\% Confidence Interval (CI) of 0.05 DMOS per model. A total of 19 teams participated in track 1 and only 3 teams participated in track 2. Figures \ref{fig:track1} and \ref{fig:track2} shows the \textit{Speech MOS}, \textit{Background Noise MOS} and the \textit{Overall MOS} results for track 1 and 2 respectively. The tables contain Differential MOS (DMOS) which is the difference between the MOS of the processed clips by a model (or the team) and the MOS of the original noisy speech. The original noisy speech MOS is shown inside the parenthesis in the row containing DMOS for "Noisy". The absolute MOS can be computed by adding the MOS of "Noisy" with the DMOS of a particular model of interest. The participants will be ranked based on \textit{Overall MOS} given they satisfy other challenge requirements. Participants are required to submit the number of operations per second of their model. This will be used as a tie-breaker. The challenge also requires the participating teams to submit a paper to INTERSPEECH 2021 explaining their method and get the paper accepted. 

\subsection{Key takeaways}
\begin{enumerate}
\item We can see from figure \ref{fig:f1a} that all the teams except for the top 2 did worse than noisy in terms of speech quality measured by \textit{Speech MOS}. Most of the noise suppressors introduce speech distortion when they get aggressive in suppressing noise and end up suppressing speech components. This is likely due to an inaccurate estimation of noise power in each frame.

\item The \textit{Background Noise MOS} in \ref{fig:f1b} shows that almost all the models are trained well to suppress the background noise. The top 3 models achieved a DMOS of almost 2 (MOS of 4.6), which is a significant improvement over noisy speech.

\item According to figure \ref{fig:f1c}, the best model achieved an \textit{Overall MOS} of about 3.78, which is about 1.2 DMOS less than the perfect quality of MOS 5. This shows that achieving superior noise suppression without distorting speech is a challenging problem. Analysis in \cite{naderi2020crowdsourcing} shows that the best theoretical \textit{Overall MOS} that can be achieved is 3.92 assuming \textit{Bacground Noise MOS} of 5 and \textit{Speech MOS} same as that of noisy speech. Hence, the field of speech enhancement optimized for human perception is still in its nascent phase.

\item Researchers can get the biggest bang for the buck in terms of \textit{Overall MOS} by improving the \textit{Speech MOS} by maintaining excellent \textit{Background Noise MOS}.

\end{enumerate}

\section{Conclusions}

The INTERSPEECH 2021 DNS Challenge was organized to help researchers from academia and industry to come together and tackle this challenging problem in speech enhancement. Large inclusive and diverse training and test datasets with supporting scripts were open sourced along with other tools such as perceptual objective metrics. Many participants from both industry and academia found the datasets very useful.

\bibliographystyle{IEEEtran}

\bibliography{mybib}


\end{document}